# Interface to Query and Visualise Definitions from a Knowledge Base

Anelia Kurteva[1]* and Hélène De Ribaupierre[2]

[1] Semantic Technology Institute, Department of Computer Science, University of Innsbruck, Austria
`anelia.kurteva@sti2.at`,
[2] School of Computer Science and Informatics, Cardiff Univeristy, Cardiff, Wales, The United Kingdom
`deribaupierreh@cardiff.ac.uk`

**Abstract.** The semantic linked data model is at the core of the Web due to its ability to model real world entities, connect them via relationships and provide context, which could help to transform data into information and information into knowledge. Linked Data, in the form of ontologies and knowledge graphs could be stored locally or could be made available to everyone online. For example, the DBpedia knowledge base, which provides global and unified access to knowledge graphs is open access. However, both access and usage of Linked Data require individuals to have expert knowledge in the field of the Semantic Web. Many of the existing solutions that are powered by Linked Data are developed for specific use cases such as building and exploring ontologies visually and are aimed at researchers with knowledge of semantic technology. The solutions that are aimed at non-experts are generic and, in most cases, information visualisation is not available. Instead, information is presented in textual format, which does not ease cognitive processes such as comprehension and could lead to problems such as information overload. In this paper, we present a web application with a user interface (UI), which combines features from applications for both experts and non-experts. The UI allows individuals with no previous knowledge of the Semantic Web to query the DBpedia knowledge base for definitions of a specific word and to view a graphical visualisation of the query results (the search keyword itself and concepts related to it).

## 1 Introduction

In this modern age of the World Wide Web, data has turned into currency and is constantly and simultaneously being exchanged between multiple entities. However, data comes in many formats. For example, textual, visual, verbal, video formats and while these formats are easily interpreted by machines and humans without having context data remains just data - a collection of facts, signals, measurements [20]. In order to become something more such as information context is needed. Adding context helps transform data into information that is organised, structured, useful and calculated [20].

---

* Corresponding author

Achieving such transformations has been the focus of the Semantic Web, which as defined in [2] is *"an extension of the current web in which information is given well-defined meaning, better enabling computers and people to work in cooperation"*. Various semantic data models such as schemas, ontologies and knowledge graphs provide the means of transforming data into knowledge by adding meaning to things through relationships [19]. In the Web of Linked Data, real world entities can be represented as concepts that are linked to each other based on a given context. Each relationship holds a specific meaning that could be interpreted both machines and humans. According to Tim Berners-Lee[1]: *"The Semantic Web is not just about putting data on the Web. It is about making links, so that a person or a machine can explore the web of data"*.

Humans are some of the main providers and consumers of data thus it is only fair that information is made available to them as well. Search engines such as Google[3], Swoogle[7], Falcons[6] are example of tools that allow one to access Linked Data. With the help of its knowledge graph[4], Google can answer one's query and further suggest relevant information in a matter of seconds. However, most of the information presented on the result's page is in textual and tabular formats, which does not ease one's comprehension and decision making. Further, the whole process is rather time-consuming when a simple question is asked. For example, a search for a word's meaning could yield hundreds of pages on Google, and as shown in [21] presenting one with too many choices could lead to confusion and loss of motivation. While machines are able to comprehend large volumes of information, written in different languages, in milliseconds this is not the case with humans. Humans are visual creatures and look for visual cues such as colors, forms, depth, and movements [5]. *"The human brain processes images 60,000 times faster than text, and 90 percent of information transmitted to the brain is visual"*[11]. Reading large volumes of information in textual formats is time-consuming and can cause problems such as information overload, which according to Gross [9] *"...occurs when the amount of input to a system exceeds its processing capacity. Decision makers have fairly limited cognitive processing capacity. Consequently, when information overload occurs, it is likely that a reduction in decision quality will occur."*.

Linked Data could be also queried directly with SPARQL[5]. This is done with the help of tools that focuses mainly on Linked Data creation and exploration such as Protégé[6] and its online version - WebProtégé[7], VOWL[8] or through application programming interfaces (APIs) such as the DBpedia REST API[9]. In order to work with Linked Data directly, for example, to query an ontology or a knowledge base such as DBpedia[10], one needs specific knowledge of the Semantic Web, namely experience with OWL[11], RDF[12] RDFs[13] and SPARQL[5]. The query results are, in most cases, still in textual format, follow a triple pattern and can include specific uniform resource identifiers (URIs), which while useful for machines, could be confusing to individuals. Software applications such as Protégé allow one to both query and explore a visual representation of a specific ontology. While this is extremely helpful for individuals with expert knowledge of the Semantic Web it is not in favour of non-experts as one needs to import a specific ontology of choice and use SPARQL to query it. Further, even if one is experienced in the field of the Semantic Web, querying DBpedia, even for a single concept, would result in displaying millions of triples thus the issue of information overload[9] arises again. The result would be confusion due to the inability to comprehend and even read the presented information thus having a visualisation of the Linked Data is simply not enough. One should consider what information is needed and how to visualise in order to ease comprehension.

---

[3] https://www.google.com
[4] https://developers.google.com/knowledge-graph
[5] https://www.w3.org/TR/rdf-sparql-query/

In this paper, we present a web application with a user interface (UI) that allows individuals with no previous knowledge of the Semantic Web to query the DBpedia knowledge base for definitions of a specific word. Further, the UI includes a graphical visualisation of the query results (i.e. the search keyword and concepts related to it). The web application combines features from applications for both experts and non-experts. It provides the options to both search and to visualise results in order to ease one's comprehension. The main research question that this paper aims to answer is:

*"Is a visualisation of a knowledge base's query result useful for individuals?"*.

Our main hypothesis is that a visualisation of the query results, when presented together with a definition of a word, is both useful and interesting to individuals.

The paper is structured as follows: Section 1 presents and introduction to the main research question. Related work is presented in Section 2, while Section 3 outlines the methodology that this research follows. Section 4 provides insights of the implemented solution. Section 5 presents the evaluation results. Conclusions and future work is presented in Section 6.

---

[6] https://Protege.stanford.edu

[7] https://webProtege.stanford.edu

[8] http://vowl.visualdataweb.org

[9] https://wiki.dbpedia.org/rest-api

[10] https://wiki.dbpedia.org

[11] https://www.w3.org/2001/sw/wiki/OWL

[12] https://www.w3.org/TR/2004/REC-rdf-primer-20040210/

[13] https://www.w3.org/2001/sw/wiki/RDFS

4      A. Kurteva, H. De Ribaupierre

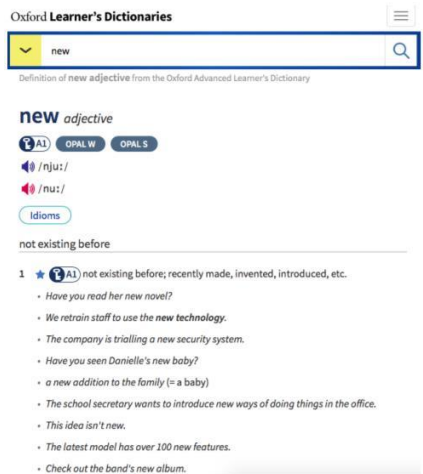 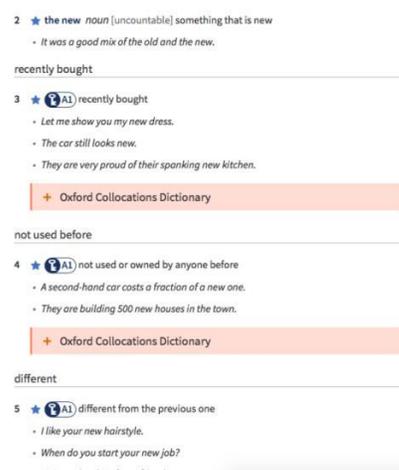

**Fig. 1.** Interface of the Oxford Learner's Dictionaries

**Fig. 2.** Results of a search query by Oxford Learner's Dictionaries

## 2   Related Work

In 2006, the creator of the World Wide Web, Tim Berners-Lee defined the following four principles[14] for sharing data on the Web: (i) use URIs as names for things, (ii) use HTTP URIs so that people can look up those names, (iii) when someone looks up a URI, provide useful information, using the standards (RDF, SPARQL) and (iv) include links to other URIs, so that more things could be discovered. Many of the existing applications that are powered by Linked Data obey these principles, however, their interfaces are either too generic or too complex. Applications aimed at professionals such as Protégé[6] provide the option to not only explore Linked Data but also to create, edit and visualise it. These options are rarely available in applications developed for non-expert. While it is true that different users have different needs regarding functionality, presenting users with visualisation of data has been proven to be useful as visualisations help ease comprehension, engage one's attention and arouse curiosity [14][15][16][25].

Online dictionaries such as Lexico[15], Oxford Learner's Dictionaries[16] and Cambridge Dictionary[17] present one with an interface, which resembles a search engine and could be used by both experts and non-experts. The main focus of these tools is on searching for word's meaning, displaying similar terms and grammar rules. While the Lexico and Oxford Learner's Dictionaries allow one to input a word and hear how it is pronounced, the Cambridge Dictionary allows one to simultaneously search for a word's meaning and its translation in different languages. Regarding the interface design, all three dictionaries resemble each other. The interface is designed as a single page view, which based on the input term is divided into several sections explaining its meaning and providing examples. Although, providing examples of how a term could be used could help one comprehend the meaning of the term better, all of the information is in textual format. Further, depending on how much information is available for a specific term, the result page could require one to scroll several times in order to get to a specific section (see Fig. 1 and Fig. 2).

---

[14] https://www.w3.org/DesignIssues/LinkedData.html
[15] https://www.lexico.com
[16] https://www.oxfordlearnersdictionaries.com
[17] https://dictionary.cambridge.org

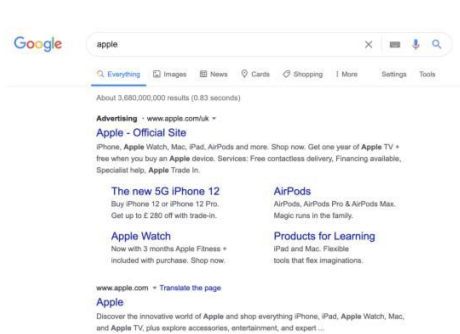 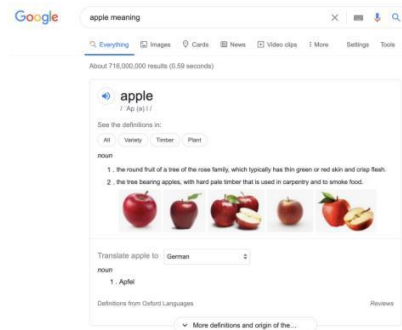

**Fig. 3.** Google search without context.    **Fig. 4.** Google search with context.

Google Search[18] is another similar tool, which is powered by a knowledge graph[8]. Google Search's capabilities exceed those of Lexico[15], Oxford Learner's Dictionaries[16] and the Cambridge Dictionary[17] as one could perform much more complicated search queries. Google's interface is simple and does not require expert knowledge in how Linked Data is structured thus it has turned into the main source of information for many. However, the results of a Google search are presented in a textual and tabular formats. Further, in some cases (Fig. 3 and 4) one needs to add specific context in order to get the desired results.

Linked Data-powered tools and applications that present information in formats other that textual are in most cases developed to be used by experts. The Protégé[6] ontology development environment is one of the go to tools for work with Linked Data as it allows one to create, import, export, query and visualise Linked Data structures such as ontologies. Protégé's interface provides the option to customise what fields are shown. Individuals can select from a variety of editing options and can further use plugins such as OntoGraph[19] to view a graphical visualisation of the current ontology. Due to its user-friendly and intuitive interface design, Protégé has become a standard for ontology development. A lightweight version of Protégé called WebProtégé[7], which could be used online, was introduced in 2008. WebProtégé's main goal is to provide a collaborative environment for ontology development that is easily accessible online. However, in order to use both Protégé and WebProtégé one needs to have knowledge of semantic models and experience with SPARQL. The main reason behind it is both tools require one to import or create an ontology first in order to explore it visually.

---

[18] https://www.google.com

[19] https://protegewiki.stanford.edu/wiki/OntoGraf

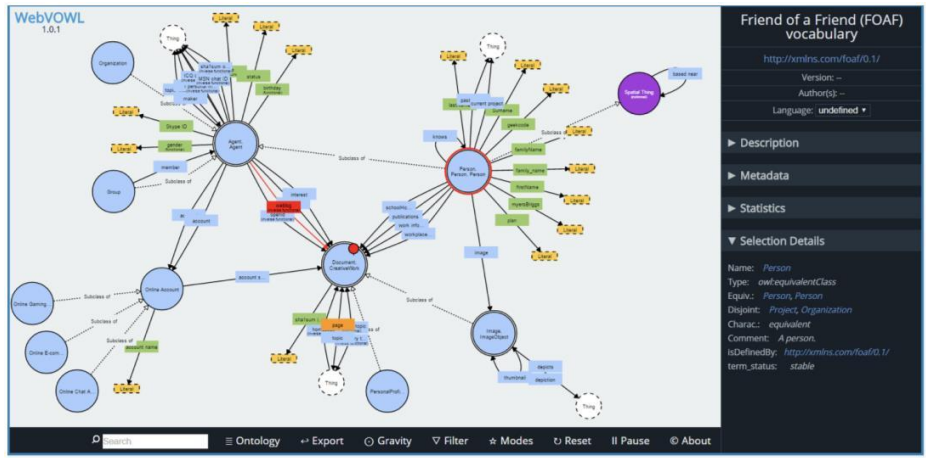

**Fig. 5.** The WebVOWL Interface

Lohman et al.[18] present VOWL[20] - an application for interactive Linked Data exploration and visualisation. VOWL is available online under the name WebVOWL[21] and as a plugin extension for Protégé[6] called ProtégéVOWL[22]. While the Protégé interface is developed with focus on easing ontology development, VOWL (Fig. 5) focuses on visualisation of Linked Data [18]. VOWL's interface allows one to import ontologies from a file or via a Uniform Resource Locator(URL). Once imported a scalable vector graphics (SVG) visualisation of the whole file is generated with D3.js[23] and is then displayed on the interface. The interface gives individuals the options to customize the generated visualisation by filtering and editing data and the colours associated with it. The evaluation of the tool, which was done with five participants (four of which were experts), showed that while helpful VOWL is *"almost showing too much information"*, which has a negative effect on one's comprehension [18].

A similarity that could be found in existing Linked Data tools for experts is their ability to generate a graphical visualisation of the data. This feature could be extremely useful when one wants to view relationships between things and discover more information. Most of the tools such as Protégé[6] and VOWL[20] require an ontology or schema to be imported or created first. Once such semantic model is available a graphical visualisation is generated for the whole ontology. A graphical visualisation of specific data could be generated only upon inputting special parameters. For example, limiting the number of results a query could return. From an experts perspective, all these options are useful but this does not apply to non-experts. However, non-experts should be given the opportunity to benefit from such visualisation. In order to achieve that, an interface that simplifies the process of Linked Data query and visualisation is needed.

---

[20] http://vowl.visualdataweb.org
[21] http://vowl.visualdataweb.org/webvowl.html
[22] http://vowl.visualdataweb.org/protegevowl.html
[23] https://d3js.org

## 3  Methodology

This work follows the methodology for building interfaces based on Linked Data presented in [17]. The methodology consists of four main steps: (i) gather data, (ii) define use case, (iii) build interface and (iv) use data. Further, we follow the recommendations for UI design by Schneiderman presented in [22]. Our main focus is on reducing one's short-term memory load (i.e. the eight rule of Schnerderman[23]) by building a one screen UI and by using hierarchical visualisations.

This work uses Linked Data that is available through the DBpedia[10] knowledge base. The main use case that we focus on is querying definitions of terms and providing a visualisation of the query results. By query results we view the definition of a word and concepts related or similar to it. Figure 6 presents this system's architecture and all of its building blocks.

A simple responsive structure for the UI was defined with HTML[24] and CSS[25]. PHP[26] and the EasyRDF[27] library were used on the server side for generating search forms and allowing SPARQL[5] queries to be executed, while JavaScript[28] was used on the client side for added interactivity. The Speak.js[29] library was used for text-to-speech transformation, which allows ones to hear a pronunciation of the search term on the UI. Regarding the visualisation itself, the D3.js[23] library was used. SPARQL was the main semantic technology used for querying the DBpedia knowledge base. The solution was hosted on a local host provided by the XAMPP[30] tool and was later made public on remote web server in order to be evaluated.

## 4  Implementation

The implementation of the proposed web application comprises of two stages. Stage 1 focuses on the UI design and implementation, while Stage 2 on the graphical visualisation of the query results. The next sections present an overview of the development at each stage.

---

[24] https://html.spec.whatwg.org
[25] https://developer.mozilla.org/en-US/docs/Web/CSS
[26] https://www.php.net
[27] https://www.easyrdf.org
[28] https://www.javascript.com
[29] https://github.com/kripken/speak.js/
[30] https://www.apachefriends.org/index.html

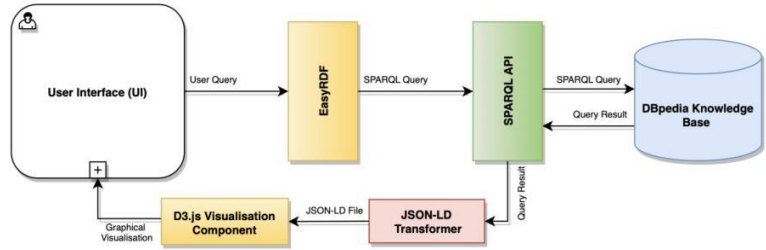

**Fig. 6.** System Architecture

### 4.1  User Interface

The main idea of the UI (Fig. 7) is to allow one to perform a keyword-based search while being as simple as possible. The UI consist of three main components: a search bar that allows the input of keywords, a "Result" field, which displays the query results in textual format and a "Visualisation" field that presents the graphical visualisation of the query results.

We focus mainly on querying definitions of words from DBpedia[10] thus we have predefined a SPARQL query that could use any word as its query variable. With the help of the EasyRDF[27] library, one's input is sent to the predefined SPARQL query, which takes it as an input variable. The query is then send to DBpedia. The main DBpedia properties we query are *"dbo:abstract"*, *"dbo:thumbnail"*, *"dbo:sameAs"*, *"rdfs:seeAlso"* and *"owl:differentFrom"*. As some concepts have longer abstracts, we limit the queried information to a few sentences in order to avoid information overload when displaying the definition to the user. All query results are stored in JSON-LD format, which allows data to be easily consumed by the D3.js[23] visualisation library. Once a query is executed and the result (i.e. the keyword, its definition and its thumbnail) is returned it is displayed in the "Result" field (Fig. 7).

When developing the UI, Human-Computer Interaction[12] was also considered. In order to try to raise one's comprehension and make one feel involved, we focused on interactivity. Components of the UI, such as the graphical visualisation, which we describe in the next section, and the fields themselves can be expanded and collapsed on demand. Further, the UI provides the option to hear a keyword's pronunciation, which is also available in the LEXICO[15] dictionary. Implementing this feature was a challenge as most of the recordings of word pronunciations were not openly available thus a simple text-to-speech functionality was implemented using the Speak.js[29] library. We made use of the *textToSpeech()* function, which allows passing a value to it and returning the specified language pronunciation.

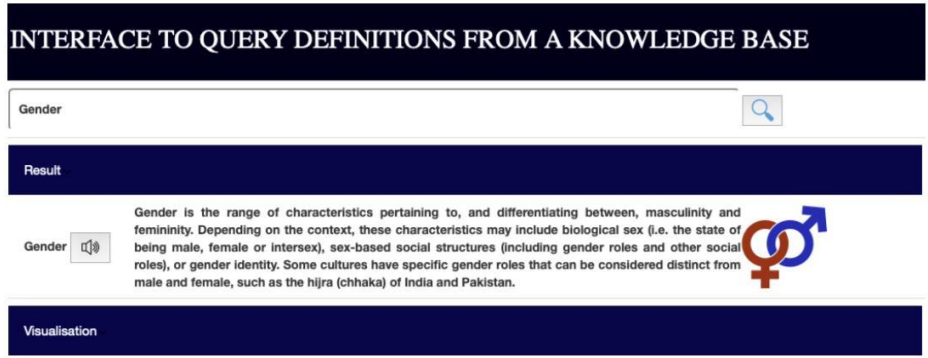

**Fig. 7.** Overview of the User Interface

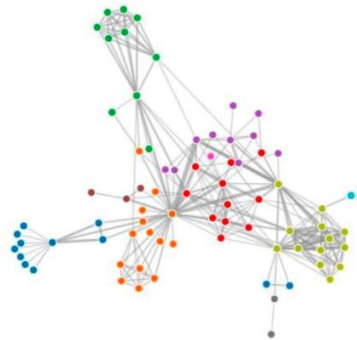 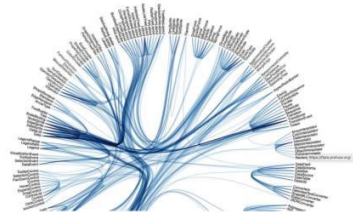

**Fig. 8.** Force-Directed Graph [3]        **Fig. 9.** Hierarchical Edge Bundling [4]

### 4.2   Graphical Visualisation of the Query Result

Linked Data visualisations provided by tools such as VOWL[18] and Protégé[6] are not aimed at users who have no previous knowledge of Graph Theory[10] and ER (Entity-Relation graph)[24] thus they limits their public to a specific group of individuals who posse such knowledge. The created visual representation (Fig. 10) of the query results for this application aims to provide a meaningful, easily understandable data representation to non-experts, who when using it feel confident and curious to learn more on their own.

In order to generate a graphical representation the D3.js[23] library was used due to its ability to handle various data formats such as JSON, JSON-LD, HTML and the vast selection of visualisation techniques and algorithms that it offers [13]. A couple of graphical representation such as a Force-Directed Graph (Fig. 8) and Hierarchical Edge Bundling (Fig. 9) exist and are widely used. The Force-Directed Graph on Fugure 8 displays all nodes and edges that are available. However, it does display information about the type relationship between the nodes when creating the edges. The visualisation itself is attractive but could cause confusion in both expert and non-expert users as no knowledge could be gained. Further, the layout is not intuitive but rather randomly scattered with no hierarchy. In comparison, the visualisation on Figure 9 follows a hierarchy. However, it would still be a challenge for individuals to understand the logic behind connecting the various terms. Based on the limitations of such graphical visualisations we have developed a visualisation algorithm that produces a graph visualisation (Fig. 10), which is both informative and visually pleasing.

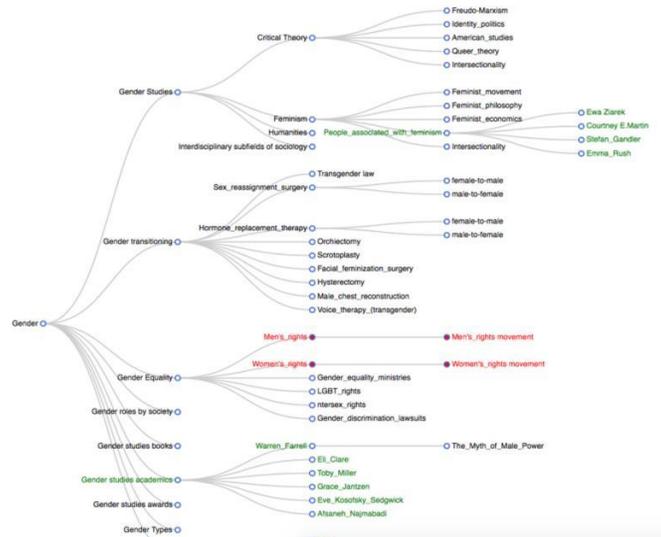

**Fig. 10.** Interactive Hierarchical Tree Layout

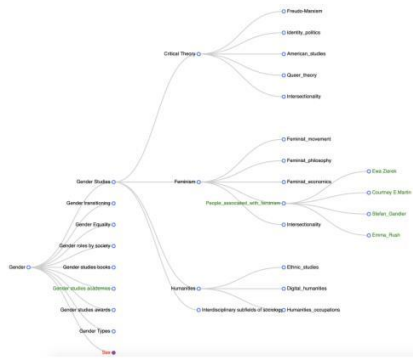         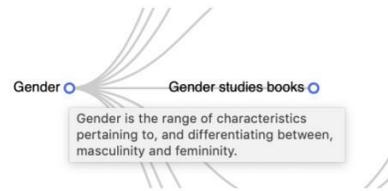

**Fig. 11.** View with Collapsed Nodes          **Fig. 12.** Tooltip with Information

The developed graph follows a hierarchical tree structure where one's input keyword could be seen as the root node. All branches coming out of it are directly connected to it based on information about the relationships queried from DBpedia[10]. Each branch represents a specific sub-category connected to the main term and has its own sub-divisions. Further, most of the external nodes are links to external sources. The main idea behind that was to awaken individuals' curiosity and allow them to continue exploring on their own. In order to differentiate between the available data types, different colours were used. The colour green was assigned to all nodes that hold information about individual personas, while the colour red was used for representing the available contradictions. All nodes that are also links to external sources change colour upon hoover. The graph is further enhanced by making each node interactive - nodes can be collapsed or expanded by clicking on them (Fig. 11) and a tooltip with information is shown upon hoovering on the root node (Fig. 12) . The options to zoom in and zoom out are available via a scroll of the mouse wheel or the equivalent touch pad action. The final version of the UI (Fig. 13) allows one to query DBpedia[10] for the meaning of a term and to view the query result in both textual and visual formats.

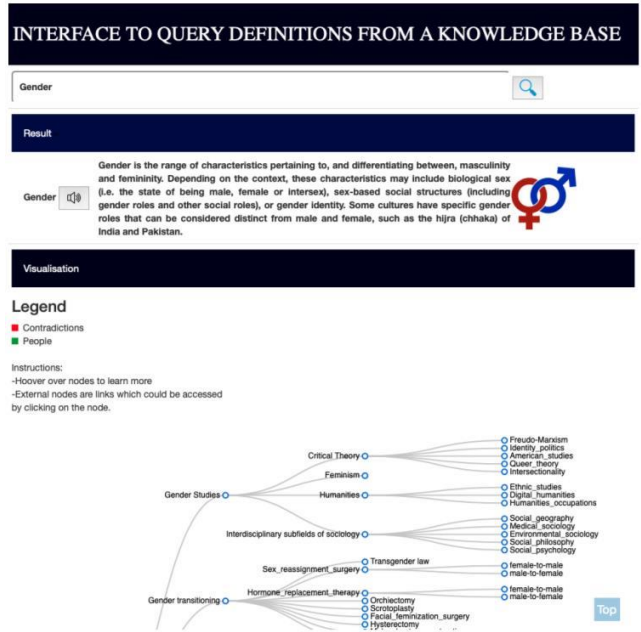

**Fig. 13.** Integrated User Interface and Graphical Visualisation

## 5   Evaluation

In order to test the usability of the developed application and its UI (Fig. 13), testing in the form of different questionnaires was conducted with 10 participants from different educational and ethnic backgrounds. The participants were presented with a system usability questionnaire, which helped evaluate the design and overall experience while using the application. The questionnaire uses a 4-point Likert scale (4-Strongly agree, 3-Agree, 2-Disagree and 1-Strongly disagree). To evaluate the usability, participants were given predefined scenarios and were asked to asked complete several tasks.

The analysis showed that 7 out of the 10 participants strongly agreed that the application was helpful and easy to use, while the rest agreed. When asked if they would use the application frequently, more than the half agreed. Regarding the visualisation itself, half strongly agreed that it was useful, while the other half gave "agree" as an answer. By computing the median values for each question, we were able to see how answers differ through the different categories. The biggest difference in answers was in the category "educational level". Participants with a Postgraduate degree strongly agreed or agreed that they would frequently use the application, while participants with both Undergraduate and High School educational level are on the fence between agree and disagree. Undergraduate users found the application not as easy to use as postgraduates and High School participants did. However, looking at the comments that all participants left on the questionnaire and while completing the given tasks, it was agreed that the application is easy to use, useful and simple. Finally, participants were given the option to describe with their own words the application and their experience. Some of the adjectives that the participants used were: useful, intuitive and accessible.

When asked about what they would improve, the participants stated: "bigger font size", "link should be easier to hoover on" and "delay in the return times". In conclusion, the evaluation showed that the application is well-accepted, user-friendly and helpful.

## 6   Conclusion

In this paper, we presented a web application for querying and visualising Linked Data aimed at non-expert users. The developed solution provides a simple, intuitive user interface, which enables users to perform different tasks such as search for definitions; interact with the graphical visualisation of their query and hear a word's pronunciation. Looking back at the main research question, we believe that the conducted evaluation provides a positive answer and proves our hypothesis as many of the patricians view the application as useful and interesting. However, there are limitations present that need to be addressed in the future.

One such limitation is the visualisation algorithm itself, which accepts as an input only a specific hierarchical JSON-LD structure. Another task for the future would be to improve the design of the graphical visualisation. This would include providing the option to select different graph layouts, optimisation of the current visualisation algorithm and of the SPARQL query. Data filters could be implemented so that a more specific result is displayed and the option to save/download the generated graph will be considered. Further, the current solu-tion allows only one-word search thus a multi-word search could be implemented as well.

In conclusion, although the presented solutions has its limitations, it achieves the task of combining features from Linked Data-powered solutions for both experts and non-experts and presents them to non-experts in an accessible way. In addition, we believe that the proposed solution in this paper could be applied in various fields. For example, it could be used as a starting point when developing artificial intelligence software tools such as virtual chatbots and next-generation search engines.

# 7    Acknowledgements

We would like to thank Simon Tippner for his helpful feedback regarding the graphical visualisation and Midhat Faheem for participating in the discussions that this research inspired.